\documentclass[12pt,fleqn]{article}
\usepackage{graphicx}
\begin{document}

\title{Non-relativistic quantum theory consistent with principle of locality }
\author{Isaac Shnaid}
\date{}
\maketitle

\begin{abstract}
Principle of locality means that any local change (perturbation) of the stationary state wave function field propagates with finite speed, and therefore reaches distant regions of the field with time delay.

If a one-particle or multi-particle non-relativistic quantum system is initially in a stationary state, and its wave function field is locally perturbed, then perturbed and non-perturbed sub-regions appear in the region. According to Schr\"odinger equation, borders of the perturbed sub-region propagate with infinite speed, and the perturbation instantaneously affects all infinite region. It means that Schr\"odinger equation predicts infinite speed of the wave function perturbations propagation. This feature of classical Schr\"odinger equation is traditionally interpreted as non-locality of quantum mechanics. From physical point of view, such mathematical behavior of Schr\"odinger equation solutions is questionable because it is difficult to accept that in real world infinite physical objects can instantaneously appear. 

We introduce a hypothesis that in reality speed of propagation of the perturbed sub-region borders is equal speed of light. On this basis we develop and analyze a finite propagation speed concept for non-relativistic quantum equations. It leads to local interpretation of non-relativistic quantum mechanics consistent with principle of locality and free of local hidden variables. The theory is applied to analysis of Einstein-Podolsky-Rosen (EPR) paradox, entanglement and properties of perturbed matter waves. We proved that formulated theory agrees with results of classical experiments on electron matter waves diffraction. 

\end{abstract}



\section{Introduction}

Principle of locality means that any local change (perturbation) of the stationary state wave function field propagates with finite speed, and therefore reaches distant regions of the field with time delay.

If a one-particle or multi-particle non-relativistic quantum system is initially in a stationary state, and its wave function field is locally perturbed, then perturbed and non-perturbed sub-regions appear in the region, where under perturbation we mean any change of wave function. According to Schr\"odinger equation, borders of the perturbed sub-region propagate with infinite speed, and the perturbation instantaneously affects all infinite region. It means that Schr\"odinger equation predicts infinite speed of the wave function perturbations propagation. This feature of classical Schr\"odinger equation is traditionally interpreted as non-locality of quantum mechanics.\footnote{The term "non-locality" is usually also applied in cases when quantum state does not depend on local  hidden variables and therefore violation of Bell's inequalities is observed in experiments.} From physical point of view, such mathematical behavior of Schr\"odinger equation solutions is questionable because it is difficult to accept that in real world infinite physical objects can instantaneously appear.

We introduce a hypothesis that in reality speed of propagation of the perturbed sub-region borders is equal speed of light. On this basis we develop and analyze a finite propagation speed concept for non-relativistic quantum equations consistent with principle of locality and free of local hidden variables. It leads to local interpretation of non-relativistic quantum mechanics. The theory is applied to analysis of Einstein-Podolsky-Rosen (EPR) paradox, entanglement and properties of matter waves. We proved that formulated theory agrees with results of classical experiments on electron matter waves diffraction. 
 
The present work is a continuation of our previous research published as e-prints \cite{sh1b} and \cite{sh1c}. 

\section{Deriving quantum equations predicting finite speed of the wave function perturbations propagation}

\subsection{Classical quantum equations and their drawbacks}

Classical time dependent Schr\"odinger equation for a multi-particle quantum system is \cite{sh8}
\begin{equation}
i \hbar \frac {\partial \Psi}{\partial t}+ \frac{\hbar^2}{2}  \sum_{i=1}^n \frac{1}{m_i}\nabla_i^2 \Psi - U \Psi=0
\label{eq:B1}
\end{equation}
where
\begin{equation}
\nabla_i^2\Psi=\frac {\partial^2 \Psi}{\partial x_i^2}+\frac {\partial^2 \Psi}{\partial y_i^2}+\frac {\partial^2 \Psi}{\partial z_i^2},
\label{eq:B2}
\end{equation} 
$\hbar=h/2\pi,~i=\sqrt{-1}$, and $h,~\Psi(G,t),~t,~U,~n,~m_i,~x_i,~y_i,~z_i$ denote Planck's constant, wave function, time, potential energy, total number of particles in the system, mass of the $i$-th particle and coordinates associated with the $i$-th particle, respectively. We define $G=\{x_1,y_1,z_1,...,x_i,y_i,z_i,....,x_n,y_n,z_n\}$ as a set of $3n$ independent coordinates of all particles in the system.

From equation (\ref{eq:B1}) follows that wave function of classical DeBroglie's  wave of a free material particle $U=0$ moving parallel axis $x_m$ is defined as \cite{sh8}, \cite{sh6}
\begin{equation}
\Psi=\exp[2 \pi i (-\nu t + k x_m)]
\label{eq:B2a}
\end{equation} 
where $\nu=E/h,~k=p/h,~E,~p$ denote frequency,  wave number, the particle energy and momentum, respectively.

Let us analyze the following thought experiment. We assume that initially the particle does not move $E=0,~p=0$ with respect to the coordinate system $x_m$. So initially there is $\nu=0,~k=0,~\Psi(x_m)=1$, and DeBroglie's wave does not exist. Then at some moment of time $t_0=0$ the particle starts moving and becomes $E \neq 0,~p \neq 0,~\nu \neq 0,~k \neq 0$, thus, creating a DeBroglie's wave and a perturbation of the initial wave function $\Psi(x_m)=1$. Here and further under perturbation we mean any local change of wave function.  In this case, according to equation (\ref{eq:B2a}),  all  infinite region instantaneously becomes perturbed, and instantaneously DeBroglie's wave (\ref{eq:B2a}) of infinite length   appears in the region.  Presented thought experiment confirms that classical time dependent Schr\"odinger equation (\ref{eq:B1}) being of parabolic type predicts infinite speed of perturbations propagation. From physical point of view, such mathematical behavior of its solutions is questionable because it is difficult to accept that in real world infinite physical objects can instantaneously appear.

According to relativistic quantum mechanics, the perturbations propagate with speed of light. However when appropriate mathematical procedures are applied to Dirac relativistic quantum equation with finite speed of the wave function perturbations propagation, only classical Schr\"odinger equation predicting infinite speed of the wave function perturbations propagation is obtained. So in non-relativistic quantum mechanics the problem persists. 

We suggest that in non-relativistic case speed of the perturbations propagation is also finite and is equal to speed of light. Further we derive non-relativistic quantum equations satisfying this principle which is equivalent to physical principle of locality. They do not contain local hidden variables and therefore predict violation of Bell's inequalities in experiments.

\subsection{Description of the wave function perturbations propagation}

According to parabolic type equation (\ref{eq:B1}), speed of the wave function perturbations propagation is infinite. To overcome this drawback, we will use method developed in our previous works \cite{sh1d}-\cite{sh3} for parabolic partial differential equations. We suggest that initially the quantum system is in a stationary state. Then at a certain moment of time $t=t_0$ wave function field is locally perturbed where under perturbation we mean any change of wave function. As a result, perturbed and non-perturbed $3n$ dimensional sub-regions appear, their boundary hyper-surface $S_P$ moves with speed $v_P$ equal speed of light into the non-perturbed sub-region, and the perturbed sub-region gradually replaces the non-perturbed one.

According to method developed in works \cite{sh1d}-\cite{sh3} we define perturbation travel-time $t_P$ as a time moment when the perturbation reached a given point $M$. Therefore the boundary hyper-surface $S_P$, separating the perturbed from the non-perturbed sub-region, is a surface of constant perturbation traveltime  $t_P=const$. For a multi-particle quantum system, $t_P,~M,~S_P$ are functions of $G$. From definition of perturbation traveltime follows that 

\begin{equation}
\sum_{i=1}^n \bigg[ \bigg ( \frac {\partial t_P}{\partial x_i}\bigg )^2+\bigg ( \frac {\partial t_P}{\partial y_i}\bigg )^2+\bigg ( \frac {\partial t_P}{\partial z_i}\bigg )^2 \bigg ]=\frac{n}{v_P^2}
\label{eq:A2}
\end{equation}

The non-linear governing equation (\ref{eq:A2}) with $3n$ independent variables is a generalization of classical eikonal equation \cite{sh4}, \cite{sh5}. Its {\em primary wave solution} $t_P(G)$ defines perturbation traveltime  field satisfying initial condition 
\begin{equation}
t_{0P}(G)=t_0
\label{eq:A3}
\end{equation}
where $t_{0P}(A)$ and $t_0$ denote initial values of traveltime and global time, respectively.

\subsection{The modified Schr\"odinger equation}

For any point $M(G)$ of the region, can be introduced local time $\vartheta(G,t)$ \cite{sh1d}-\cite{sh3} counted from a moment when the perturbation reached this point   
\begin{equation}
\vartheta(G,t)=t-t_P(G)
\label{eq:A4}
\end{equation}

Three characteristic cases are:
\begin{enumerate}
\item $\vartheta=t-t_P(G)<0$, the perturbation has not reached the point $M(G)$. Thus, the point belongs to the non-perturbed subregion.
\item $\vartheta=t-t_P(G)=0$, the perturbation has reached the point $M(G)$. Now the point $M(G)$ is located on the border hyper-surface $S_P(G)$, separating the perturbed and the non-perturbed subregions.
\item $\vartheta=t-t_P(G)>0$, the point $M(G)$ is located inside the finite perturbed subregion with moving border $S_P(G)$. All further analysis is related to this subregion where $\vartheta > 0$. Obviously, classical Schr\"odinger equation for multi-particle systems (\ref{eq:B1}) does not describe such situation, and a new governing equation is needed.
\end{enumerate}

According to the local time concept formulated earlier in works \cite{sh1d}-\cite{sh3}, parabolic type partial differential equations predicting finite speed  of the perturbations propagation and the respective equations with infinite speed  of the perturbations propagation are identical, if  local time $\vartheta=t-t_P(G)$ is used as an independent variable instead of global time $t$. When speed of the perturbations propagation is infinite $v_P \rightarrow \infty$, local time in these equations is identical to global time $\vartheta=t$ because $ t_P=0 $. The same equations describe the finite $v_P$ case  when $t_P > 0,~ 0 < \vartheta < t$. All equations written in such universal form we call \emph {modified  equations}.

Application of local time concept leads to the following formulation of modified time dependent Schr\"odinger equation for multi-particle systems

\begin{equation}
i \hbar \frac {\partial \Psi(\vartheta,G)}{\partial \vartheta}+ \frac{\hbar^2}{2}  \sum_{i=1}^n \frac{1}{m_i}\nabla_i^2 \Psi(\vartheta,G) - U \Psi(\vartheta,G)=0
\label{eq:A5}
\end{equation}

\section{Basic properties of the modified Schr\"odinger equation}

Modified Schr\"odinger equation (\ref{eq:A5}) is of the parabolic type as classical Schr\"odinger equation (\ref{eq:B1}). In a classical case of infinite speed of the perturbations propagation $v_P \rightarrow \infty$ there is $t_P=0,~\vartheta=t$, and modified equation (\ref{eq:A5}) becomes identical to classical equation (\ref{eq:B1}).

It is easy to prove that modified Schr\"odinger equation (\ref{eq:A5}) predicts finite speed of the perturbations propagation. Solutions of classical time dependent Schr\"odinger parabolic type equation (\ref{eq:B1}) suggest that any local perturbation introduced at initial  moment of time $t_0$, instantaneously affects all infinite space domain. The modified time dependent Schr\"odinger parabolic type equation (\ref{eq:A5}) uses instead of global time $t$ local time $\vartheta$  as an independent variable. In this case, any local perturbation introduced at initial local time moment $\vartheta_0=t_0-t_{0P}=0$,  affects an arbitrary point $M(G)$ of the perturbed sub-region border at the same local time value $\vartheta_M=t_M-t_P(M)=0$.  Therefore the perturbation arrives to the point $M(G)$ at global time moment $t_M=t_P(M)>0$, i.e. with global time delay. So the modified Schr\"odinger equation (\ref{eq:A5}) predicts finite speed of the perturbations propagation equal to speed of light, and the perturbations have a wave behavior. These waves do not reflect and interfere, because only solutions of eikonal  type equation (\ref{eq:A2}) corresponding to {\em  primary perturbation waves} propagating in the non-perturbed subregion, have physical meaning. 

Let  $\Psi(G,t)$ be a solution of the classical  Schr\"odinger equation (\ref{eq:B1}), satisfying certain boundary and initial conditions. For the same boundary and initial conditions, a solution of the modified Schr\"odinger equation (\ref{eq:A5}) is the same function but with local time as an argument $\Psi(G,\vartheta),~\vartheta>0$. For an arbitrary point $M(G)$, located inside the perturbed sub-region, and corresponding perturbation traveltime value $t_P(M)$ difference of solutions of classical and modified  Schr\"odinger equations is defined by the following approximate formula
\begin{equation}
\Delta \Psi=\Psi(G,t)-\Psi(G,\vartheta=t-t_P) \approx \frac {\partial \Psi}{\partial t} ~ t_P
\label{eq:A6}
\end{equation}

Absolute value of this difference is small for small values of perturbation traveltime $t_P$ and/or slow processes, when absolute value of $\frac {\partial \Psi}{\partial t}$ is small. In this case, solutions of classical time dependent Schr\"odinger equation for multi-particle systems are accurate enough. 

For a case of constant energy of the quantum system $E=const$, potential energy $U$ not depending on time, and infinite speed of the perturbations propagation, wave function can be presented as $\Psi=\psi(G) \exp (-i \frac {E}{\hbar} t)$ \cite{sh8}, \cite{sh6}, \cite{sh7}. From the local time concept follows that for finite speed of the perturbations propagation the right formula is  $\Psi=\psi(G) \exp (-i \frac {E}{\hbar}  \vartheta)$, and we  obtain time independent version of modified Schr\"odinger equation (\ref{eq:A5}) describing \emph{stationary state of multi-particle quantum system}

\begin{equation}
\frac{\hbar^2}{2}  \sum_{i=1}^n \frac{1}{m_i}\nabla_i^2 \psi +(E - U) \psi=0
\label{eq:A7}
\end{equation}

Obviously, time independent version of modified Schr\"odinger equation (\ref{eq:A7}) for multi-particle systems is identical to classical time independent Schr\"odinger equation following from expression (\ref{eq:B1}). Therefore all solutions of modified time independent Schr\"odinger equation are identical to solutions of classical equation.  The only difference is: in the classical case multiplier $\exp (-i \frac {E}{\hbar} t)$ is a function of global time $t$, while in the case of modified Schr\"odinger equation multiplier $\exp (-i \frac {E}{\hbar}  \vartheta)$ is a function of local time $\vartheta$. Therefore the steady state solution exists inside a finite perturbed space domain with a moving boundary $\vartheta=0$.

\section{Quantum interaction between the perturbed and non-perturbed sub-regions}

Let $\Psi_0(\vartheta_0,G),~\Psi_p(\vartheta_P,G),~\vartheta_0=t-t_{P0},~\vartheta_P=t-t_P,~t_{P0},~t_P$ be wave function for the non-perturbed sub-region, wave function for the perturbed sub-region, local time for the non-perturbed sub-region defined during its prehistory, local time for the perturbed sub-region, traveltime for the non-perturbed sub-region defined during its prehistory, and traveltime for the perturbed sub-region, respectively.

The following two modified Schr\"odinger equations describe quantum states of the sub-regions

\begin{equation}
i \hbar \frac {\partial \Psi_0(\vartheta_0,G)}{\partial \vartheta_0}+ \frac{\hbar^2}{2}  \sum_{i=1}^n \frac{1}{m_i}\nabla_i^2 \Psi_0(\vartheta_0,G) - U \Psi_0(\vartheta_0,G)=0
\label{eq:A7a}
\end{equation}

\begin{equation}
i \hbar \frac {\partial \Psi_P(\vartheta_P,G)}{\partial \vartheta_P}+ \frac{\hbar^2}{2}  \sum_{i=1}^n \frac{1}{m_i}\nabla_i^2 \Psi_P(\vartheta_P,G) - U \Psi_P(\vartheta_P,G)=0
\label{eq:A7b}
\end{equation}

As the non-perturbed subregion is in stationary state and the perturbed sub-region is in non-stationary state, we can present the wave functions $\Psi_0$ and $\Psi_P$ in the following forms

\begin{equation}
\Psi_0(\vartheta_0,G)=\psi_0(A)~exp\bigg(-i\frac{E_0}{\hbar} \vartheta_0\bigg)
\label{eq:A7c}
\end{equation}

\begin{equation}
\Psi_P(\vartheta_P,G)=\psi_P(\vartheta_P,G)~exp\bigg(-i\frac{E_0}{\hbar} \vartheta_0\bigg)
\label{eq:A7d}
\end{equation}
where $E_0$ is total energy of the non-perturbed multi-particle system.

Now expression (\ref{eq:A7c}) is substituted in equation (\ref{eq:A7a}), and expression (\ref{eq:A7d}) in (\ref{eq:A7b}). As $\frac{\partial \vartheta_0}{\partial \vartheta_P}=1$, we obtain

\begin{equation}
\frac{\hbar^2}{2}  \sum_{i=1}^n \frac{1}{m_i}\nabla_i^2 \psi_0 +(E_0 - U) \psi_0=0
\label{eq:A7e}
\end{equation}

\begin{equation}
i \hbar \frac {\partial \psi_P}{\partial \vartheta_P}+ \frac{\hbar^2}{2}  \sum_{i=1}^n \frac{1}{m_i}\nabla_i^2 \psi_P + (E_0-U)\psi_P=0
\label{eq:A7f}
\end{equation}

Equation (\ref{eq:A7f}) can be solved using initial condition $\psi_P(0,G)=\psi_0(G)$ obtained from equation (\ref{eq:A7e}) and an additional function $\psi_P(\vartheta_P,G_b)$ describing local perturbation of wave function at the point $M(G_b)$. The solution of equation (\ref{eq:A7f}) $\psi_P(\vartheta_P,G)$ and formula (\ref{eq:A7d}) give wave function of the perturbed sub-region $\Psi_P(\vartheta_P,G)$. Thus, becomes clear why the perturbed sub-region gradually replaces non-perturbed sub-region and not vice versa.

If the non-perturbed sub-region is in non-stationary state, presentation (\ref{eq:A7e}) and (\ref{eq:A7f}) becomes impossible. It means that perturbed sub-region does not appear, and in this case the perturbations propagation is non-observable.

In a particular classical case, quantum state of every particle is fully defined, i.e. wave function of every particle $\Psi_i(t,x_i,y_i,z_i)$ is known. Therefore wave function of multi-particle system is a product of wave functions of all particles \cite{sh8}

\begin{equation}
\Psi(t,G)=\prod_{i=1}^n \Psi_i(t,x_i,y_i,z_i)
\label{eq:A7g}
\end{equation}

Expression (\ref{eq:A7g}) assumes infinite speed of the wave function perturbations propagation. For finite speed of the wave function perturbations propagation, according to the local time concept the following formulas determine wave functions in the non-perturbed and perturbed sub-regions 

\begin{equation}
\Psi_0(\vartheta_0,G)=\prod_{i=1}^n \Psi_{0i}(\vartheta_0,x_i,y_i,z_i)
\label{eq:A7h}
\end{equation}

\begin{equation}
\Psi_P(\vartheta_P,G)=\prod_{i=1}^n \Psi_{Pi}(\vartheta_P,x_i,y_i,z_i)
\label{eq:A7i}
\end{equation}
where $\Psi_{0i},~\Psi_{Pi}$ is non-perturbed and perturbed wave function of $i$-th particle, respectively.

When the distance between particles is big enough and $t_0=0$, then for $i$-th particle approximate formula holds $\vartheta_P=t-d_i/v_P$ where $d_i$ denotes distance between the particle, where initial perturbation was introduced, and  $i$-th particle. Obviously, as a result of finite speed of the wave function perturbations propagation, at the same local time the bigger global time delay $\Delta_i t=d_i/v_P$ will be observed for distant particles with bigger $d_i$.

\section{Local interpretation of EPR paradox and entanglement}

EPR paradox \cite{sh9}, \cite{sh9a} reflects one of the most peculiar and principal features of classical non-relativistic quantum mechanics. The authors A. Einstein, B. Podolsky and N. Rosen considered two quantum systems I and II that initially interacted and then interaction stopped. Now certain quantity is measured in the system I. It appears that according to classical non-relativistic quantum  mechanics, the measurement in the system I instantaneously affects state of the system II, isolated from the system I. If then another quantity is measured in system I, another state of the isolated system II instantaneously appears. Thus, classical quantum theory suggests paradoxical non-local superluminal action on distance. The authors wrote: "We are thus forced to conclude that the quantum-mechanical description of physical reality given by wave function is not complete" \cite{sh9}. Today this conclusion is not supported by physical community \cite{sh9b}, though the community does not reject the EPR paradox proof presented in \cite{sh9}.

Obviously, EPR paradox arises as a result of infinite speed of the wave function perturbations propagation suggested by classical non-relativistic quantum mechanics. The EPR paradox can be solved using developed concept of the wave function perturbations propagation with finite speed. After initial interaction, systems I and II create a two-particle system which is in stationary state described by equation (\ref{eq:A7e}). Any measurement in the system I is a local perturbation of the two-particle system wave function field. This perturbation propagates to system II with speed of light according to eikonal type equation (\ref{eq:A2}), while perturbed field is described by equation (\ref{eq:A7f}).

If quantum state of every particle is fully defined, then local description of EPR paradox give expressions (\ref{eq:A7h}) and (\ref{eq:A7i}) and done above appropriate analysis based on them.

Thus, developed local theory predicts, that as a result:
\begin{enumerate}
\item Quantum state of the system II will change when the wave function perturbation will arrive. 
\item This change will occur with time delay.
\item Described interaction of the systems I and II has local character and is transferred with speed of light. 
\end{enumerate}

The same way of reasoning is applicable to any multi-particle system which is left in a stationary state after initial interaction of the particles. If then quantum state of one of the particles  is perturbed, the perturbation wave according to equation (\ref{eq:A2}) travels with speed of light to other particles affecting their state, thus, creating entanglement. 

Developed theory does not use any hidden variables. Because of this it predicts that in experiments will be observed violation of the Bell's inequalities - a feature fully confirmed by numerous experiments.

\section{Perturbed waves of matter according to modified Schr\"odinger equation}

From modified Schr\"odinger equation (\ref{eq:A5}) follows that perturbed wave function of  DeBroglie's matter wave of a free material particle is 
\begin{equation}
\Psi=\exp[2 \pi i (-\nu \vartheta + k x_m)]
\label{eq:B3}
\end{equation}
and $k=p/h$ is wave number of classical matter wave (\ref{eq:B2a}).

We suggest that initial local perturbation was introduced at $t=0,~x_m=0$, and the particle moment $p$ and speed $v$ are positive. It means that the particle moves in positive direction of $x_m$ axis. In the perturbed region $v_P t>x_m>-v_P t$, whose boundaries move with speed of light, two perturbation waves exist. Wave $A_m$ propagates in the same direction, as the particle, and occupies region $v_P t>x_m>0$. Wave $B_m$ moves in opposite direction in the region $0>x_m>-v_P t$. For these waves traveltime is defined as
\begin{equation}
t_P=\xi x_m \frac {1}{v_P}
\label{eq:B4}
\end{equation}
where $\xi=1$ for wave $A_m$, and $\xi=-1$ for wave $B_m$.

As $\vartheta=t-t_P=t-\xi x_m \frac {1}{v_P}$, expression (\ref{eq:B3}) becomes
\begin{equation}
\Psi=\exp[2 \pi i (-\nu t + k_l x_m)]
\label{eq:B5}
\end{equation}
where $k_l$ is wave number of the perturbed matter wave
\begin{equation}
k_l=k+\xi \frac{\nu}{v_P}
\label{eq:B6}
\end{equation}

Formula (\ref{eq:B6}) proves that for wave $A_m$, where $\xi=1$, wave number $k_l$ is bigger than classical wave number $k$, and wavelength $\lambda_l=1/k_l$ is smaller than classical wavelength $\lambda=1/k$. 

From expression (\ref{eq:B6}) also follows that wave $B_m$ with $\xi=-1$ has wave number $k_l$ lower than classical wave number $k$ and wavelength $\lambda_l=1/k_l$ higher than classical wavelength $\lambda=1/k$. 

Let $v_{ph.l}=\nu/k_l,~v_{gr.l}=d\nu/dk_l,~v_{ph}=\nu/k,~v_{gr}=d\nu/dk$ be modified wave phase velocity, modified wave group velocity, classical wave phase velocity and classical wave group velocity, respectively. From expression (\ref{eq:B6}) we obtain

\begin{equation}
\frac{1}{v_{ph.l}}=\frac{1}{v_{ph}}+\frac{\xi}{v_P}
\label{eq:B7}
\end{equation}
\begin{equation}
\frac{1}{v_{gr.l}}=\frac{1}{v_{gr}}+\frac{\xi}{v_P}
\label{eq:B8}
\end{equation}

As in non-relativistic case phase and group velocities are $v_{ph}=v/2$ and $v_{gr}=v$ \cite{sh8}, \cite{sh6}, \cite{sh7}, we obtain

\begin{equation}
v_{ph.l}=\frac{v}{2+\xi v/v_P}
\label{eq:B9}
\end{equation}
\begin{equation}
v_{gr.l}= \frac{v}{1+\xi v/v_P}
\label{eq:B10}
\end{equation}
Formulas (\ref{eq:B7})-(\ref{eq:B10}) show that phase and group velocities of perturbed waves $A_m$ and $B_m$ depend on sign of $\xi$ and always are positive.

\section{Experimental check of the developed theory}

Classical experiments of C.J. Davisson \cite{sh10} and G.P. Thomson \cite{sh10a}-\cite{sh12} and their colleagues on electron matter waves diffraction were repeated many times and respective literature is enormous, however we could not find more detailed and reliable experimental data than published by these pioneers. This is the reason why we use their results for experimental check of our theoretical suggestions.

\subsection{Davisson-Germer experiments}

Nobel lecture of C.J. Davisson \cite{sh10} includes results of classical Davisson and Germer experiments with electron waves diffraction on a crystal of nickel.  The lecture presents experimental values of electron wave length $\lambda_{exp}$ vs. $V^{-0.5}$, where $V$ is electron accelerating voltage. These data can be used for experimental check of formula (\ref{eq:B6}) following from our theory because $k_{exp}=1/\lambda_{exp}$ and $v=(2 e V/m)^{0.5}$, where $e,~k_{exp}$ denote electric charge of electron and experimental value of electron wave number, respectively.

Fig. 1 shows experimental points (diamonds) with values of $k_{exp}$ and $v$ corresponding to respective values  of $\lambda_{exp}$ and $V$ in lecture \cite{sh10}. Line A is classical $v_P \rightarrow \infty$ DeBroglie's electron wave number $k=m v/h$ vs. electron speed $v$. Though a part of experimental points are close enough to A $k_{exp} \approx k$, for most of experimental points there is significantly $k_{exp} > k$, and are missing points where substantially $k_{exp} < k$.

\begin{figure}
\center
\includegraphics[width=12cm, height=10cm]{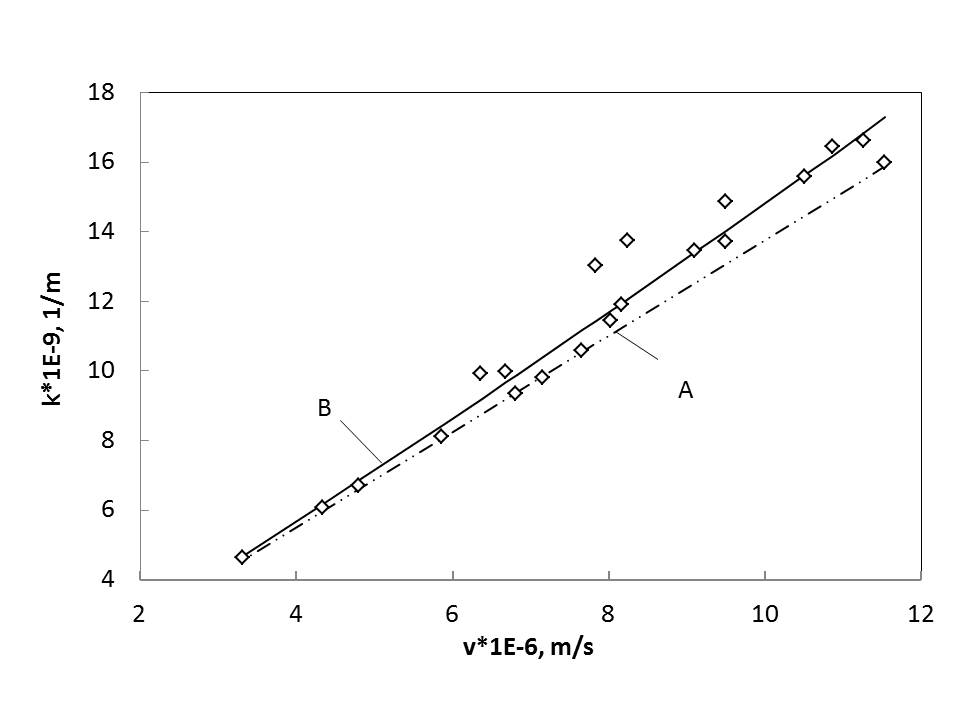}
\caption{Electron wave number vs. electron speed in Davisson and Jermer experiment (diamonds - experimental data from \cite{sh10}; A - theoretical curve for classical DeBroglie's wave with infinite speed of the perturbations propagation; B - best fit curve for perturbed DeBroglie's wave with finite speed of the perturbations propagation).}
\end{figure}

Line B in Fig. 1 is best fit (least  squares) approximation for all experimental points, obtained by variation of $v_P$ in formula (\ref{eq:B6}) for modified wave number $k_l$ and $\xi=1$. For line B variance is $4.9*10^{17}~ 1/m^2$, and for line A variance is significantly bigger $11.2*10^{17}~ 1/m^2$. Line B corresponds to speed of the perturbations propagation equal $v_P=1.3*10^8$ m/s, which is 44\% of speed of light. Author of paper \cite{sh10} does not present experimental errors. Because of this, we are not able to find why experimental value of speed of the perturbations propagation differs from our theoretical prediction. Most probably it is the result of experimental errors. Anyway obtained experimental value of speed of the wave function perturbations propagation has order of magnitude of speed of light.

So analyzed experimental data 
\begin{description}
\item confirms correctness of formula (\ref{eq:B6}) for modified wave number; 
\item supports hypothesis of finite speed of the wave function perturbations propagation, and supplies its experimental value of $v_P=1.3*10^8$ m/s, which is 44\% of speed of light;
\item proves that modified Schr\"odinger equation (\ref{eq:A5}) is physically sound.
\end{description}

\subsection{Thomson experiments}

Now we will analyze experiments with electron matter waves diffraction by thin poly-crystalline films \cite{sh10a}-\cite{sh12} done by G.P. Thomson and his colleagues. The authors used  Hull-Debye-Sherrer method initially developed for X-rays diffraction analyses of thin poly-crystalline films. They sent an approximately homogeneous beam of accelerated electrons through a very thin poly-crystalline film at normal incidence and studied a diffraction pattern registered by a photographic plate located at some distance from the film. The pattern is a family of concentric rings with diameters $D_\mu=a_\mu/k$, where $\mu=1,2,3,...$ is a number of the ring starting from the smallest $D_\mu$; $a_\mu$ denotes a coefficient, depending on material of the film, number $\mu$ of the diffraction ring and distance from the film to the photographic plate; while $k$ is wave number of electron matter wave. Therefore there is $a_\mu=D_\mu k$, and for the same poly-crystalline film, the same distance from the film to the photographic plate, and the same $\mu$, the product $D_\mu k$ does not depend on electron accelerating voltage $V$ affecting $k$. Authors of cited works \cite{sh10a}-\cite{sh12} found experimentally that the product $D_\mu k$ does not vary significantly if $k$ is calculated using classical DeBroglie's formula. They concluded that classical quantum mechanics is correct, but we will show that our theory better agrees with their experimental data. 

\begin{table}[h]
\caption{Main results of calculations using experimental data ($c$ is speed of light)}
\vspace*{5mm} 
\centering
     \begin{tabular}
     {| l | c | c | c | c | c | r |}
     \hline
     &&&&&&\\
    Crystalline material, $\mu$& Data source &$r_{var}$& $v_P/c$ & $\overline {a}_{\mu l} \cdot 10^{-9}$ &$\overline {a}_\mu \cdot 10^{-9}$& $\xi$\\ 
     &&&&&&\\
     \hline     
     Celluloid, $\mu=1$&\cite{sh11}&0.58&2.16&1.43&1.54&-1\\
     \hline
     Gold, $\mu=1$&\cite{sh11}&0.26&1.59&2.65&3.02&-1\\
     \hline
     Aluminium, $\mu=1$&\cite{sh11}&0.21&1.39&2.66&3.11&-1\\
     \hline
     Platinum, $\mu=1$&\cite{sh11}&0.43&1.07&2.72&3.05&-1\\
     \hline
     Silver, $\mu=1$&\cite{sh12}&0.79&1.63&2.28&2.63&-1\\
     \hline
     Tin, $\mu=1$&\cite{sh12}&0.46&1.07&2.69&2.25&1\\
     \hline
     Copper, $\mu=3$&\cite{sh12}&0.38&1.37&2.20&2.60&-1\\
     \hline
     Copper, $\mu=4$&\cite{sh12}&0.45&2.97&3.24&3.49&-1\\
     \hline
     Copper, $\mu=5$&\cite{sh12}&0.06&1.47&3.59&4.18&-1\\
     \hline
     Copper, $\mu=6$&\cite{sh12}&0.06&1.05&3.95&4.99&-1\\
     \hline
     Copper, $\mu=7$&\cite{sh12}&0.42&0.83&7.66&5.99&1\\
     \hline     
     \end{tabular}
 \end{table}

Expression (\ref{eq:B6}) following from our theory was applied to experimental data presented in cited works \cite{sh10a}-\cite{sh12}. For given material of the film, and for every value of $V$ and $\mu$  we calculated $E,~ p,~k,~a_\mu=D_\mu k,~k_l,~a_{\mu l}=D_\mu k_l$. Then for every set of $V$, corresponding to certain film material, the same distance of photographic plate from the film and $\mu=const$, we determined mean values $\overline{a}_\mu,~\overline{a}_{\mu l}$ and respective variances of experimental data  $var_\mu,~var_{\mu l}$  corresponding to infinite and finite speed of the perturbations propagation $v_P$, respectively. As variance $var_{\mu l}$ depends on $v_P$, we take into account only such value of $v_P$ that ensures minimal value of variance and therefore minimal ratio  $r_{var}=var_{\mu l}/var_\mu$. Main results of these calculations are presented in the Table 1. They do not include data from the source \cite{sh10a} because presented there $V$ values are not accurate as the author claims in his next paper \cite{sh11}. The table also does not present data for celluloid film from \cite{sh11b}, aluminium film in short camera from \cite{sh11} and lead film from \cite{sh11a}, because all these data are not accurate enough. Experimental results for platinum film in short camera \cite{sh11} include only two points, while for copper film \cite{sh12} only three values of $V$ for $D_1$ and $D_2$ are given, that is not enough for accurate conclusions, and these data are also not introduced in the table.

The table shows that in all analyzed cases $r_{var}<1$. It means  that hypothesis of finite speed of the wave function perturbations propagation in non-relativistic quantum mechanics and its corollary modified Schr\"odinger equation (\ref{eq:A5}) are confirmed experimentally. Mean value of dimensionless speed of the perturbations propagation calculated using data from the table and data for the Davisson-Germer experiments is $v_P/c=1.4$ with standard error of mean value equal 0.2. So 95\% confidence limits for dimensionless speed of the perturbations propagation are $1.0 \leq v_P/c \leq 1.8$. Therefore suggested in theory value $v_P/c=1$ is within these 95\% confidence limits. Thus, the cited experiments support our hypothesis that speed of the perturbations propagation is equal the speed of light. 

\section{Conclusions}

\begin{enumerate}
\item The hypothesis of finite speed of the wave function perturbations propagation in non-relativistic quantum mechanics is introduced and analyzed. The hypothesis includes assumption that speed of the perturbations propagation is equal speed of light
\item Formulated eikonal type equation defining perturbation traveltime and modified Schr\"odinger equation create theoretical basis for analyses of non-relativistic quantum systems. The developed theory is  consistent with physical principle of locality and is free of local hidden variables thus ensuring violation of Bell's inequalities in experiments. 
\item Formulated theory gives local interpretation of the Einstein-Podolsky-Rosen paradox and entanglement of quantum particles. 
\item Developed theory predicts that real characteristics of DeBroglie's matter waves are different than predicted by classical theory.
\item Classical experiments on electron matter waves diffraction by mono-crystals (Davisson-Germer experiment) and thin poly-crystalline films (Thomson experiment) support developed theory.
\end{enumerate}

\end{document}